\documentstyle[12pt]{article}
  \newcommand{\be}{\begin{equation}}
  \newcommand{\ee}{\end{equation}}
  \newcommand{\bea}{\begin{eqnarray}}
  \newcommand{\eea}{\end{eqnarray}}
  \newcommand{\lb}{\label}
\renewcommand{\a}{\alpha}

\renewcommand{\d}{\delta}
  
\renewcommand{\l}{\lambda}
\renewcommand{\L}{\Lambda}
  
\renewcommand{\S}{\Sigma}

  \newcommand{\bcal}[1]{{\cal #1}}
  \newcommand{\calb}{\bcal{B}}
  \newcommand{\ad}{\mbox{ad}}
  \newcommand{\Ad}{\mbox{Ad}}
\renewcommand{\Re}{\mbox{Re}}
\renewcommand{\Im}{\mbox{Im}}
  \newcommand{\ldel}{\langle}
  \newcommand{\rdel}{\rangle}
  \newcommand{\sprod}[2]{\ldel\,#1\,,\,#2\,\rdel}
  \newcommand{\lbrak}[2]{[#1,#2]}
  \newcommand{\Min}{\mathop{\rm Min}}
  \newcommand{\supp}{\mathop{\rm supp}}
  \newcommand{\bra}[1]{\ldel\,#1\,|}
  \newcommand{\ket}[1]{|\,#1\,\rdel}
  \newcommand{\braket}[2]{\ldel\,#1\,|\,#2\,\rdel}

\begin{document}
%
\begin{titlepage}
\begin{flushright}
ZU-TH 4/94
\end{flushright}
\vspace{3 ex}
\begin{center}
\vfill
{\LARGE\bf Instability of Einstein-Yang-Mills Solitons for arbitrary
Gauge Groups}
\vfill
{\bf Othmar Brodbeck and Norbert Straumann}
\vskip 0.5cm
Institute for Theoretical Physics\\University of Z\"urich\\
Winterthurerstrasse 190, CH-8057 Z\"urich
\end{center}
\vfill
\begin{quote}
We prove that static, spherically symmetric, asymptotically flat,
regular solutions of the Einstein-Yang-Mills equations are unstable for
arbitrary gauge groups. The proof involves the following main steps.
First, we show that the frequency spectrum of a class of radial
perturbations is determined by a coupled system of radial
``Schr\"odinger equations". Eigenstates with negative eigenvalues
correspond to exponentially growing modes. Using the variational
principle for the ground state it is then proven that there always
exist unstable modes (at least for ``generic" solitons). This
conclusion is reached without explicit knowledge of the possible
equilibrium solutions.
\end{quote}
\vfill
\end{titlepage}
\section{Introduction}
In recent years the study of regular and black hole solutions of the
Einstein-Yang-Mills (EYM) equations, without and with additional fields
(Higgs fields, dilaton fields, etc.) has been actively pursued, and a
number of
interesting and surprising results have been discovered. Among the
regular solutions the most interesting ones are those for which gravity
is
essential. The first example of this type was found by Bartnik and
McKinnon \cite{bartnik} for the $SU(2)\,$-EYM system. For the same
model several authors \cite{volkov,bizon,kunzle0} discovered later the
colored black hole solutions which showed that the classical uniqueness
theorem for the Abelian case does not generalize. The existence of both
types of solutions, which meanwhile has been established
rigorously \cite{smoller1,smoller2,smoller3}, led to a search for
corresponding solutions of other related field theories. It turned out,
for instance, that the Einstein-Skyrme system has black hole solutions
with hair which are at least linearly stable
\cite{droz,heusler1,heusler2,heusler3}. (For a numerical investigation
of nonlinear stability, see
ref. \cite{heusler3}.) Several authors studied other models, notably
the $SU(2)\,$-EYM system with a Higgs triplett
\cite{lee,ortiz,breitenlohner,aichelburg}, as well as the EYM-dilaton
system \cite{lavrelashvili}, and found in some cases other linearly
stable black hole
solutions. Interesting black hole solutions have recently been found
numerically for the EYM system with a Higgs doublet \cite{greene}, as
in the standard electroweak model. These ``sphaleron black holes" are
expected to be unstable, but this question is not yet fully analyzed.
We have shown recently, that the regular solutions are unstable
\cite{boschung}, but our method cannot be applied directly to the black
hole case.

The Bartnik-McKinnon solution and the related black hole solutions were
shown to be unstable by some of us \cite{NS1,NS2,zhou1,zhou2}. On
physical grounds one would expect that this remains true for any gauge
group, but a mathematical
proof of this conjecture presents quite a challenge. We finally
succeeded in constructing such a proof for the regular solutions, which
will be sketched in the present paper. Similar arguments may perhaps
also be used successfully to establish the instability of all black
hole solutions, but problems related to the boundary conditions at the
horizon have not yet been overcome.

Our strategy is based on the study of the pulsation equations
describing linear radial perturbations of the equilibrium solutions.
Because the derivation of these equations relies heavily on our
previous work \cite{OB1,OB2}, we first recall some basic facts and
equations in the next section. The pulsation
equations are then presented in section 3 in a convenient partially
decoupled form. It turns out, that the frequency spectrum of a class of
radial perturbations is determined by a coupled system of radial
``Schr\"odinger equations" whose bound states correspond to
exponentially growing modes. In section 4 we prove with the variational
principle for the ground state that
there always exist unstable modes, at least in the generic case
(defined in sect.\ 2). The construction of the trial variations, which
are used to establish this, is presumably the main point of the present
letter; we succeeded only after a number of failures.
\section{Basic facts and equations}
Since we consider only spherically symmetric configurations, the metric
$g$ of the space-time manifold $M$ can be parametrized as
\be
g= -NS^2 dt^2 + N^{-1} dr^2 + r^2({d\vartheta}^2 +
\sin^2\!\vartheta\,{d\varphi}^2)\:,
\lb{ns1}
\ee
where the metric functions $N$ and $S$ depend only on the
Schwarzschild-like radius $r$ and the time $t$. We use also the usual
mass fraction $m(r,t)$, defined by $N=:1-2m/r$. Gauge fields with
spherical symmetry have been described in detail in ref.\ \cite{OB1}.
Let us briefly recall the results, as far as they will be needed
below.

We fix a maximal torus $T$ of $G$ with the corresponding integral
lattice $I$ (= kernel of the exponential map restricted to the Lie
algebra $LT$ of the torus). In addition, we choose a basis $S$ of the
root system $R$ of real roots, which defines the fundamental Weyl
chamber
\be
K(S)=\{\, H\in LT \mid \alpha (H)>0 \,\mbox{ \rm for all } \alpha\in S
\,\}\:. \lb{ns2}
\ee

To a given spherically symmetric configuration, there belongs a
canonical element $H_\lambda\in I\cap\overline{K(S)}$ which
characterizes the corresponding principal bundle $P(M,G)$ admitting an
$SU(2)$ action (see sect.\ II of ref.\ \cite{OB1}). In \cite{OB2} we
showed that  $H_\lambda$ is restricted to a small finite subset of
$I\cap\overline{K(S)}$, if the solution is regular at the origin. In
much of our discussion we exclude the possibility that $H_\lambda$ lies
on a boundary of the fundamental Weyl chamber and consider only what we
call the ``generic case", for which $H_\lambda$ is in the {\em\/open\/}
Weyl chamber $K(S)$.

A spherically symmetric gauge field can be described by a smooth family
of linear maps $\L\colon LSU(2)\rightarrow LG$, depending only on $r$
and $t$ and satisfying the following conditions:
\be
\L_1=[\L_2,\L_3]\:, \qquad
\L_2=[\L_3,\L_1]\:, \qquad
\L_3 =-H_\l/4\pi\:,
\lb{ns3}
\ee
where $\L_k:=\L(\tau_k)$ and $2i\tau_k$ are the Pauli matrices. These
equations imply that $\L_+:=\L_1+i\L_2$ lies in the following direct
sum of root spaces $L_\alpha$:
\be
\renewcommand{\arraystretch}{1.3}
\begin{array}{cc}
{\displaystyle
\Lambda_+ \in  \bigoplus_{\alpha\in\S}\;L_\a}\:,
\\
{\displaystyle
\S:=\{\,\a\in R_+\mid\a(H_\l)=2\,\}}\:.
\lb{ns4}
\end{array}
\renewcommand{\arraystretch}{1}
\ee
In the generic case $\S$ is a basis of a root system contained in $R$
(see Appendix A of ref.\ \cite{OB2}).

For the example of the gauge group $SU(2)$, $H_\l$ is an integer
multiple of $4\pi\,\tau_3$: $H_\l=4\pi k\,\tau_3$ with $k\in{\cal Z}$,
and the only solutions of (\ref{ns3}) are $\L_1=\L_2=0$,
$\L_3=k\,\tau_3$, or
\be
\L_1=w\,\tau_1+\tilde{w}\,\tau_2\:,\qquad
\L_2=\mp\tilde{w}\,\tau_1\pm w\,\tau_2\:,\qquad
\L_3=\pm\tau_3\:.
\lb{ns5}
\ee

In ref.\ \cite{OB1} it is shown that there exists always a (local)
gauge such that the gauge potential $A$ takes the form
\be
A=\tilde{A}+\hat{A}\:,
\lb{ns6}
\ee
with
\be
\hat{A}={\Lambda_2
\,d\vartheta}+{(\Lambda_3\cos\vartheta-\Lambda_1\sin\vartheta)\,d\varphi}
\lb{ns7}
\ee
and
\be
\tilde{A}=NS{\cal A}\,dt+{\cal B}\,dr\:,
\lb{ns8}
\ee
where ${\cal A}$ and ${\cal B}$ commute with $H_\l$ (i.e. with $\L_3$).
If $H_\l$ is generic one knows that its centralizer is the
infinitesimal torus $LT$. Hence, ${\cal A}$ and ${\cal B}$ are
$LT$-valued and $\tilde{A}$ is thus Abelian.

The coupled EYM equations, corresponding to the parametrizations
(\ref{ns1}), (\ref{ns6}) -- (\ref{ns8}) have been derived in
\cite{OB1}. Here, it suffices to write them in a slightly different
form for the temporal gauge $\bcal{A}=0$.

The Einstein equations reduce to ($\kappa:=8\pi G$)
\be
m'= {\kappa\over 2}
\Bigl\{
NG+p_\theta
\Bigr\}\:,
\qquad
\dot m={\kappa\over 2}NH\:,
\lb{ns9}
\ee
\be
{S'\over S}={\kappa\over r}G\:,
\lb{ns10}
\ee
where
\addtolength{\jot}{5pt}
\begin{eqnarray}
G &=& \frac{1}{2}
\Bigl\{
(NS){\vphantom{|\L_+|}}^{-2}\,|\dot\L_+|{\vphantom{|\L_+|}}^2
	  +|\,\L_{+}'+[\bcal{B},\L_+]\,|{\vphantom{|\L_+|}}^2
\Bigr\}\:,\lb{ns11} \\
H &=& \mbox{Re}\,
\bigl\ldel\,
\dot\L_{+}\,,\,\L_{+}'+[\bcal{B},\L_+]\,
\bigr\rdel\:,\lb{ns12}\\
p_\theta &=& {1\over 2r^{2}}
\Bigl\{
|\check {\cal F}|{\vphantom{|{\cal F}}}^2
 +|\hat {\cal F}|{\vphantom{|{\cal F}}}^2
\Bigr\}\lb{ns13}
\end{eqnarray}
\addtolength{\jot}{-5pt}
with
\be
\check{\cal F} = \frac{r^2}{S}\,\dot\bcal{B}\:,
\qquad
\hat  {\cal F} = \frac{i}{2}\,[\L_+,\L_-]-\L_3\:.
\lb{ns14}
\ee
Here we have used the following notation: $\ldel\,\cdot\,,\cdot\,\rdel$
denotes a suitably normalized $\Ad (G)$-invariant scalar product on
$LG$, as well as its (hermetian) extension to $LG_{\cal C}$, and
$|\cdot|$ is the corresponding norm.

The YM equations decompose into

\vspace{.5ex}
\be
\frac{2}{NS}
\biggl(
\frac{r^2}{S}\,\dot\bcal{B}
\biggr)\!{\vphantom{\biggr)}}^{\textstyle\cdot}
+ \Bigl[\,\L_+\,,\,\L_-'+[\bcal{B},\L_-]\,\Bigr]
+ \Bigl[\,\L_-\,,\,\L_+'+[\bcal{B},\L_+]\,\Bigr]=0\:,
\lb{ns15}
\ee

\vspace{-2.25ex}
\begin{eqnarray}
\frac{1}{S}
\biggl(
\frac{1}{NS}\,\dot\L_+
\biggr)\!{\vphantom{\biggr)}}^{\textstyle\cdot}
\!\!\!&-&\!\!\!
\frac{1}{S}
\Bigl(NS\Bigl\{\,
\L_+'+[\bcal{B},\L_+]
\,\Bigr\}\Bigr)'
\nonumber\\
\!\!\!&-&\!\!\!
N
\Bigl\{\,
[\bcal{B},\L_+']+\big[\,\bcal{B}\,,\,[\bcal{B},\L_+]\,\big]
\,\Bigr\}
+\frac{i}{r^2}\,[\hat {\cal F},\L_+]=0\:,
\lb{ns16}
\end{eqnarray}

\vspace{-1.5ex}
\be
2
\biggl(
\frac{r^2}{S}\,\dot\bcal{B}
{\biggr)}'
+ 2\frac{r^2}{S}[\bcal{B},\dot\bcal{B}\,]
+\frac{1}{NS}
\Bigl\{\,
[\L_+,\dot\L_-]+[\L_-,\dot\L_+]
\,\Bigr\}
=0\:.
\lb{ns17}
\ee

\vspace{3ex}\noindent
The last equation is the Gauss constraint. For the generic case the
term proportional to $[\bcal{B},\dot\bcal{B}\,]$ in (\ref{ns17})
vanishes.

For static solutions all time derivatives disappear and the basic
equations simplify considerably. (For the Bartnik-McKinnon solution
$\L$ is of the form (\ref{ns5}) with $\tilde w=0$, $\L_3=\tau_3$ and
$\tilde A=0$ in (\ref{ns6}).)
\section{Pulsation equations}
We consider now a static, regular, asymptotically flat solution of the
coupled EYM equations (\ref{ns9}), (\ref{ns10}), (\ref{ns15}) --
(\ref{ns17}) with a generic $H_\l$. Such a solution is of purely
magnetic type with vanishing YM charge. This is not yet rigorously
proven under satisfactory weak assumptions, but there is strong
evidence for this (see \cite{OB2,kunzle} for partial results). From now
on the symbols $\L_\pm$, $N$, $S$, etc. will be used to denote the
equilibrium solution. Time-dependent perturbations are denoted by
$\d\L_\pm,\d\bcal{B}$, etc..

The details of the reduction of the perturbation equations to a
well-adapted form will be described elsewhere. Here, we emphasize only
some crucial points and present the final result.

An observation which was already pointed out in \cite{NS1} for
$G=SU(2)$ finds here a natural generalization which simplifies matters
considerably: The first of the Einstein equations (\ref{ns9}) leads to
an expression for $\d m'$ which can be readily integrated with respect
to $r$. An arbitrary function of $t$ which thereby appears for $\d m$
is then fixed by the second Einstein equation in (\ref{ns9}). We find
\be
\d m=\frac{\kappa}{2}N\,\Re\,\sprod{\L_+'}{\d\L_+}\:.
\lb{ns19}
\ee

It turns out that a significant decoupling of the perturbation
equations is achieved by decomposing $\d\L_+$ as follows. Let us choose
in the space (\ref{ns4}) a basis $\{ e_\a\}$ of the root spaces $L_\a$
($\a\in\S$), with respect to which we expand the unperturbed $\L_+$ and
its perturbation $\d\L_+$,
\be
\d\L_+=\sum_{\a\in\S}\,\d w^\a\, e_\a\:.
\lb{ns20}
\ee
Then we have ($\d\L_-:=c\,(\d\L_+)$, $c=$ conjugation in $LG_\bcal{C}$)
\be
\d\L_\pm=\d X_\pm \pm i\d Y_\pm
\lb{ns21}
\ee
with
\be
\d X_+=\sum_{\a\in\S}\,\Re\,(\d w^\a)\: e_\a\:,\qquad
\d Y_+=\sum_{\a\in\S}\,\Im\,(\d w^\a)\: e_\a\:.
\lb{ns22}
\ee

We shall call $\d X_\pm$, $\d Y_\pm$ the real and imaginary parts of
the perturbations $\d\L_\pm$. The unperturbed quantity $\L_+$ can be
chosen to have only a real part, as was shown in \cite{OB2}.

With the help of the equilibrium equations one can now bring the
perturbation equations after some work into the following standard
form: $\d X_+$ decouples,
\be
\d \ddot X_+ +\,U_{XX}\,\d X_+=0\:,
\lb{ns23}
\ee
and $\d\Phi :=(\d Y_+,\d\bcal{B})$ satisfies
\be
T\d \ddot\Phi + U\d\Phi=0\:,
\lb{ns24}
\ee
where the operators $U_{XX}$ and $U$,
\be
\renewcommand{\arraystretch}{1.5}
U=\left( \begin{array}{cc}
     U_{YY} & U_{Y\calb} \\
U_{\calb Y} & U_{\calb\calb}
\end{array} \right)\:,
\lb{ns25}
\renewcommand{\arraystretch}{1}
\ee
are given by the expressions
\addtolength{\jot}{8pt}
\begin{eqnarray}
U_{XX}&=&
{p_\ast}^2\,+\,\frac{NS^2}{r^2}\ad(i\hat\bcal{F})
\,-\,\frac{1}{Nr^2}\,\ad(NS\L_+)\,\ad(NS\L_-)\nonumber\\
&&{}\:-\:(\,p_\ast\L_+)\:\frac{\kappa}{r}
\biggl\{
\frac{(NS)'}{NS}+\frac{1}{r}
\biggr\}
\sprod{\,p_\ast\L_+}{\cdot\,}\nonumber\\
&&{}\,+\:(\,p_\ast\L_+)
\:\frac{\kappa S}{r^3}
\sprod{\,\lbrak{\hat\bcal{F}}{\L_+}}{\cdot\,}
\:+\:\lbrak{\hat\bcal{F}}{\L_+}
\:\frac{\kappa S}{r^3}
\sprod{\,p_\ast\L_+}{\cdot\,}\lb{ns26}\:,\\
U_{YY}&=&{p_\ast}^2\,+\,\frac{NS^2}{r^2}\ad(i\hat\bcal{F})\:,\lb{ns27}\\
U_{\calb\calb}&=& -\:\ad(NS\L_+)\,\ad(NS\L_-)\:,\lb{ns29}\\
U_{Y\calb}&=&
+\:p_\ast\,\ad(NS\L_+)\,+\,\ad(NS\,p_\ast\L_+)\:,\lb{ns28}\\
U_{\calb Y}&=& -\:\ad(NS\L_-)\,p_\ast\,+\,\ad(NS\,p_\ast\L_-)
\end{eqnarray}
\addtolength{\jot}{-8pt}
with
\be
p_\ast=-iNS\frac{d}{dr}\;.
\lb{ns30}
\ee

The operator $T$ in (\ref{ns24}) acts as multiplication with the
diagonal matrix
\be
\renewcommand{\arraystretch}{1.5}
T=\left( \begin{array}{cc}
1 & 0 \\
0 & Nr^2
\end{array} \right)\:.
\lb{ns31}
\renewcommand{\arraystretch}{1}
\ee

It is easy to see that $U$ and $T$ are symmetric relative to the
following scalar product for $LG_{\bcal{C}}$-valued functions on
$\bcal{R}_+$:
\be
\braket{\Psi}{\Phi}=\int_0^\infty\sprod{\Psi}{\Phi}\;\,\frac{dr}{NS}\;.
\lb{ns32}
\ee
(Questions of domains of unbounded operators will be discussed
elsewhere.)

For a harmonic time dependence, $\d \Phi=\xi\mbox{e}^{-i\omega t}$, we
obtain from (\ref{ns24})
\be
{\omega}^2 =\: \frac{\bra{\xi}\:U\,\ket{\xi}}
		      {\bra{\xi}\:T\,\ket{\xi}}\:.
\lb{ns33}
\ee
Different eigenmodes are orthogonal with respect to the following
scalar product defined by $T$:
$\braket{\xi_1}{\xi_2}_T:=\bra{\xi_1}\:T\,\ket{\xi_2}$. For the
frequency $\omega_0$ of the fundamental mode of eq. (\ref{ns24}) we
have the minimum principle
\be
{\omega_0}^2 =\Min_\xi\: \frac{\bra{\xi}\:U\,\ket{\xi}}
			      {\bra{\xi}\:T\,\ket{\xi}}\:,
\lb{ns34}
\ee
which will be used in the next section to show that ${\omega_0}^2<0$.

The perturbation equations (\ref{ns23}), (\ref{ns24}) do not include
the Gauss constraint (\ref{ns17}), whose linearization reads
\be
\frac{d}{dt}
\Bigl\{\,
p_\ast\biggl(\frac{r^2}{S}\d\calb\biggr)-\lbrak{\L_+}{\d Y_-}
\,\Bigl\}
=0\:.
\lb{ns35}
\ee
It will turn out that this is automatically satisfied for physical
pulsations (more precisely: in the space orthogonal to pure gauge
modes).
\section{Instability of all generic EYM solitons}
For a given regular solution with $\L_+=\sum_{\a\in\S}\, w^\a e_\a$ we
construct now a one-parameter family of field configurations
$\L_{(\chi)+}$, ${\calb}_{(\chi)}$ such that the variational
expressions (\ref{ns34}) for
$\d\L_\pm=(d\L_{(\chi)\pm}/d\chi)_{\chi=0}$,
$\d\calb=(d{\calb}_{(\chi)}/d\chi)_{\chi=0}$ is {\em\/ negative \/}.
This family is chosen of the following form:
\bea
\L_{(\chi)+}&=&\Ad(\exp(-\chi Z))
\Bigl\{\,
\L_+\cos(\chi)+i\L_+(\infty)\sin(\chi)
\,\Bigr\} \lb{ns36}\:,\\
\calb_{(\chi)}&=&\chi Z'\:,
\lb{ns37}
\eea
where $Z$ is an $LT$-valued function of $r$ with the properties
\be
\lim_{r\to 0,\infty}\:\lbrak{Z}{\L_+}=i\L_+(\infty)\:,\qquad
\supp Z'\subseteq[\,1-\epsilon\,,1+\epsilon\,]
\lb{ns38}
\ee
for an $\epsilon >0$. The existence of such a function can be seen as
follows. Let $\{h_\a\}_{\a\in\S}$ be the dual basis of $2\pi\S$ and put
\be
Z=\sum_{\a\in\S}\, Z^\a h_\a\:,\qquad
Z^\a  =  \left \{
\begin{array}{ll}
w^\a (\infty)/w^\a (0)   & \mbox{ for }r<1-\epsilon\:,\\
1                        & \mbox{ for }r>1+\epsilon\:.
\end{array} \right.
\lb{ns39}
\ee
It is easy to verify that both conditions in (\ref{ns38}) are
satisfied. (It can be shown that $w^\a (0)\neq 0 \mbox{ for all }
\a\in\S$; see Appendix A of ref.\ \cite{OB2}.)

We note a few properties of the family (\ref{ns36}), (\ref{ns37}). The
equilibrium solution is clearly obtained for $\chi=0$. Applying a gauge
transformation with $g=\exp(\chi Z)$ we obtain
\be
\L_{(\chi)+}\to\L_+\cos(\chi)+i\L_+(\infty)\sin(\chi)\:,\qquad
\calb_{(\chi)}\to 0\:.
\lb{ns40}
\ee
The first variations of (\ref{ns36}) and (\ref{ns37}) are
\be
\d\L_+=-\lbrak{Z}{\L_+}+i\L_+(\infty)\:,\qquad
\d\calb=Z'\:,
\lb{ns41}
\ee
and these satisfy by construction the desired boundary conditions
\be
\lim_{r\to 0,\infty}\d\L_+=0\:,\qquad
\lim_{r\to 0,\infty}\d\calb=0\:.
\lb{ns42}
\ee
($\d\calb$ vanishes even outside $[\,1-\epsilon\,,1+\epsilon\,]$.)
Finally, $\d\L_+$ has only an imaginary component
\be
\d Y_+=\d\L_+=-\lbrak{Z}{\L_+}+i\L_+(\infty)
\lb{ns43}
\ee
and thus by (\ref{ns34})
\be
{\omega_0}^2 \leq\: \frac{\bra{\d\Phi}\:U\,\ket{\d\Phi}}
		    {\bra{\d\Phi}\:T\,\ket{\d\Phi}}\:.
\lb{ns44}
\ee
with $\d\Phi=(\d Y_+,\d\calb)$ given by (\ref{ns43}) and the second
eq.\ in (\ref{ns41}).

It is easy to see that the denominator in this expression is finite if
$\L_+(r)-\L_+(\infty)$ vanishes sufficiently fast for $r\to\infty$.

The calculation of the numerator in (\ref{ns44}) for the operator $U$
in (\ref{ns25}) is somewhat tedious. After several partial
integrations, using (\ref{ns42}) and the equilibrium equations, one
finally arrives at
\be
\bra{\d\Phi}\:U\,\ket{\d\Phi}=
-\int
\:\Bigl\{
 NS\,{|\L_+'|}^2+2\frac{S}{r^2}|\hat\bcal{F}|{\vphantom{|\L_+'|}}^2
\,\Bigr\}\,dr\;,
\lb{ns45}
\ee
which is finite and strictly negative!

We now return to the linearized Gauss constraint (\ref{ns35}) and note
that a variation $\d\Phi$ is orthogonal with respect to
${\braket{\,\cdot\,}{\,\cdot\,}}_T$ to all gauge variations
\be
\d\Phi_{\mbox{gauge}}=(\,i\lbrak{\chi}{\L_+}\,,\chi'\,)\:,\qquad
\chi\colon \bcal{R}_+\to LT\:,
\ee
if and only if the curly bracket in (\ref{ns35}) vanishes. This can be
shown readily with a partial integration.

This proves the instability of the regular equilibrium solution. It
should be emphasized that we were able to draw this conclusion assuming
only weak asymptotic conditions for the solitons. Especially, we have
used repeatedly that the YM charge vanishes only if
$\lim_{r\to\infty}\L(r)$ is a homomorphism from $LSU(2)$ to $LG$ (see
eq.\ (\ref{ns14})).
\section*{Acknowledgments}
We would like to thank Markus Heusler for discussions and comments and
to Michael Volkov for suggestive remarks on stability problems. This
work was supported in part by the Swiss National Science Foundation.
\end{document}